\documentstyle{mn}

\title[ Relativistic Effect of Gravitational Deflection
        of Light in Binary Pulsars]
     { Relativistic Effect of Gravitational Deflection
       of Light in Binary Pulsars}
\author[O.V.Doroshenko \& S.M.Kopeikin]
       {O.V.Doroshenko$^1$\thanks{On leave from
           Astro Space Centre of P.N.Lebedev
           Physical Institute, Leninskii prospect 53,
	   Moscow 117924, Russia}
       and S.M.Kopeikin$^2$\\
       $^1$Max-Planck-Institut f\"ur Radioastronomie,
           Auf dem H\"ugel 69, D-53121 Bonn, Germany\\
       $^2$Laboratory of Astronomy and Geophysics,
           Hitotsubashi University, Kunitachi, Tokyo 187, Japan
       }
\date{Accepted 5 January 1995. Received November 1994.}

\pagerange{\pageref{firstpage}--\pageref{lastpage}}
\pubyear{1995}

\begin{document}

\label{firstpage}

\maketitle

\begin{abstract}
An improved formula for the timing of binary pulsars that accounts for the
relativistic deflection of light in the gravitational field of the
pulsar's companion is presented, and the measurability of this effect
together with its variance estimates are discussed. The deflection of the
pulsar's beam trajectory in the gravitational field of its companion leads
to variation in the pulsar's rotational phase. This variation appears as the
narrow sharp growth of the magnitude of the post-fit pulsar timing residuals
in the vicinity of the moment of the superior conjunction of the pulsar with
its companion. In contrast to the relativistic Shapiro effect
the amplitude of the effect of gravitational deflection of the
pulsar radio beam has two peaks
with opposite signs, which become sharper as the inclination $i$ of
the pulsar's orbit approaches 90$^{\circ }$. The effect under consideration
influences the estimation of parameters of the relativistic Shapiro effect
in the binary pulsars with nearly edgewise orbits. Its inclusion in the
fitting procedure provide more careful measurement of sine of the orbital
inclination $i$ as well as the masses of the pulsar $m_p$ and its companion
$m_c$. This permits an improved testing of alternative theories of gravity
in the strong field
regime. Moreover, the measurement of this effect
permits independent geometric constraints
on the position of the pulsar
rotational axis in space. The effect of the gravitational deflection of light
has been numerically investigated for binary pulsars with nearly
edgewise orbits. It is
shown that the effect is observed in general only when $\cos i\leq 0.003.$
This estimate becomes less restrictive as the pulsar's spin axis approaches
the line of the sight.
\end{abstract}


\section{Introduction}

High-precision timing observations of binary pulsars on relativistic orbits
provide unique opportunities for testing of General Relativity (GR) through
two mechanisms: the emission of gravitational radiation and the presence of
strong-field effects. The effect of emission of gravitational waves have
been confirmed using the PSR B1913+16 long term data with an accuracy better
than 0.5\% (Taylor \& Weisberg 1989; Damour \&Taylor 1991). This was
accomplished by \
the measurements of three relativistic effects: 1) the rate of apsidal
motion $\dot \omega $, 2) the Doppler and gravitational red shifts $\gamma $
of pulsar rotational frequency, and 3) the system orbital period decay $\dot
P_b$, which is due to the effect of the gravitational waves emission, which has
been
rigorously justified from the mathematical point of view by Damour
(1983a,1983b) and, independently, by Grishchuk \& Kopeikin (1983,1986), as
well as Sch\"afer (1985). However, timing of PSR\ B1913+16 is not sufficiently
sensitive to the strong gravitational field effects. And, indeed, the
theoretical
investigations of Damour and Esposito-Far\`ese (1992) have shown that some
alternative to the GR theories of gravity, which include the additional
scalar fields, also fit the above mentioned $\dot \omega
-\gamma -\dot P_b$ test, and, hence, can not be distinguished from the GR
theory. Realizing this fact has led to the development of the parameterized
post-Keplerian (PPK) formalism, which was designed to extract the maximum of
possible information from the high-precise observational data for binary
pulsars (Damour \& Taylor 1992). The application of this methodology to the
four pulsars B1913+16, B1855+09, B1534+12 and B1953+29 has already
ruled out a number of the alternative gravity theories (Taylor {\em et al.}
1992).

The PPK formalism is especially effective in discriminating alternative
theories of gravity in the strong field regime when there is a possibility
of independently measuring two post-Keplerian parameters chaacterizing the
range $r\equiv Gm_c/c^3$ and the shape $s\equiv \sin i$ of the famous
Shapiro delay in propagation of radio signals of pulsar in the gravitational
field of its companion. The most favorable case for the measurements of $r$
and $s$ is when the pulsar's orbit is inclined to the plane of the sky at
the angle $i$ close to 90$^{\circ}$. Timing observations of the
''non-relativistic'' binary pulsar B1855+09 by Ryba \& Taylor (1991), and by
Kaspi {\it et al.} (1993) have provided the very first such independent
measurements of the parameters $r$ and $s$ that allowed the determination of
the
neutron star mass with a precision better than about 18\%. It is worthwhile to
stress
that such mass determination does not require a knowledge of the rate of
relativistic apsidal motion $\dot \omega $ and the red shift parameter $%
\gamma$ as it takes place in the case of the PSR B1913+16 system.

At the same time when a binary pulsar orbit is viewed nearly edge-on (as for
B1855+09 and PSR B1534+12) the complementary effect in the propagation of
the pulsar's radio signal may be important. This is due to the well known
effect of the deflection of light rays in the gravitational field of a
self-gravitating body (in the gravitational field of the pulsar's companion
in the event of the binary pulsar system). Due to the gravitational
deviation of the pulsar's beam from a straight line, the observer will see
the pulsar's pulse only when the pulsar is rotated by an angle compensating
this deviation. In contrast to the relativistic Shapiro effect, the effect
of gravitational deflection of light directly contributes to the rotational
phase of the pulsar (as, for instance, the effect of aberration of pulsar's
beam
does), but this is so small that it can be misinterpreted as an additional
delay in
the time-of-arrivals (TOA) of the pulsar pulses. Similar to the Shapiro
delay, the deflection of the pulsar's beam by the gravitational field of its
companion manifests itself in the timing as the rapid, sharp growth of the
magnitude of the post-fit residuals of TOA on a short time interval in the
vicinity to the moment of superior conjunction of the pulsar and its
companion. The effect in question is superimposed on the Shapiro effect, and,
in principle, should be explicitly taken into account to avoid an incorrect
determination of the parameters $r$ and $s$.

Another important feature of the effect of gravitational deflection of light
in a binary pulsar is a strong dependence of the shape of TOA residuals on
the spatial orientation of the pulsar's rotation axis (see the eqs. (7), (9),
(10) in the text below). Thus, the measurement of this effect would help to get
further constrains on the position of the pulsar's spin independently
of the polarization observations. Let us emphasize that the effect of the
relativistic aberration caused by the orbital motion of pulsar in binary
systems (Smarr \& Blandford 1976) also strongly depends on the spin's
orientation, but can not be observed directly due to the same
time-dependence on the pulsar's orbital phase as the classical Roemer
effect (Damour \& Deruelle 1986).

Epstein (1977) was, probably, the first who called attention to the
phenomenon of deflection of the pulsar radio beam in the gravitational field of
its companion when $\cos i\leq 10^{-3}$. However, he did not make any
calculations confirming this estimate. Narayan {\it et al. }(1991) also
mentioned this effect as a useful tool for investigation of black hole
- neutron star binary systems and verification of the existence of the black
hole to an unparalleled degree of certainty. Schneider (1989, 1990)
calculated the corresponding gravitational deflection time delay and
lensing amplification factor of the pulse intensity in the edge-on binary
pulsar,
but, unfortunately, his presentation is not easily followed.

In the present paper we try to solve the problem more simply, giving both a
complete theoretical treatment and a discussion of the measurability of the
effect of the gravitational deflection of radio beam in binary pulsars
which we call briefly the bending delay (BD) effect. It is shown how to
include this effect in the fitting procedure of TOA for the evaluation of
the pulsar parameters of binary pulsars whose orbital inclinations to the
plane of the sky are extremely close to 90$^{\circ }$ .

In Sec. 2 of the paper we present an improved formula for the timing of binary
pulsars with the BD effect taken into account. In Sec. 3 we investigate,
using numerical simulations, the measurability of the masses
and orbital inclination in binary systems when the BD effect is taken into
account. A mathematical procedure for the direct measurement of parameters of
the BD effect is outlined in Sec. 4 and applied to process the observational
data for PSR B1855+09. In Sec. 5 we present a summary of our results and
conclusions.

\section{Timing formula}

The self-consistent mathematical derivation of a timing formula for the
calculation of pulsar rotational phase should be based upon the relativistic
theory of astronomical reference systems together with the matching
procedure developed previously by Kopeikin (1988), Brumberg \& Kopeikin
(1989a,b) and Brumberg (1991). The exact timing
formula must include the kinematic effects of the galactic differential
rotation, orbital and precessional motions of the pulsar, relativistic
effect of aberration, quadratic Doppler and red shift effects, the Shapiro
delay, as well as the deflection of the pulsar beam in the gravitational
field of its companion. In the present paper we ignore the effects of
differential galactic acceleration and precessional motion of the pulsar
rotational axis suggesting for simplicity that vectors of the angular
velocity $\vec \Omega $ and spin $\vec S$ of the pulsar are coincident. The
timing formula obtained is a slightly modified version of that elaborated
by Damour \& Deruelle (1986). Some additional theoretical aspects of
the derivation of the timing formula are discussed by Klioner \& Kopeikin
(1994) and Kopeikin (1994, 1995).

Let $\eta $ and $\lambda $ be the polar coordinates of the pulsar's spin
axis $\vec S$, where $\eta $ is the longitude of the pulsar's spin axis in
the plane of the sky measured from the ascending node of the pulsar's
orbital plane ($0^{\circ }\leq \eta <360^{\circ })$, and $\lambda $ is the
angle between the pulsar's spin axis and the line of sight pointing from the
terrestrial observer toward the pulsar ($0^{\circ }\leq \lambda <180^{\circ
})$. The time evolution of the pulsar rotational phase $\phi (T)$ is given
by the equation:
\begin{equation}
\label{1}\phi(T)/2\pi\equiv N(T)=N_0+\nu_pT+\frac12\dot \nu_pT^2+
\frac16\ddot\nu_pT^3\ ,
\end{equation}
where $N_0$ is the initial phase; $\nu_p,\dot \nu_p,\ddot \nu_p$ denote the
pulsar spin frequency and its time derivatives shifted by the Doppler
factor, and $T$ is the pulsar proper time related to the solar-system
barycentric arrival time $t$ according
to the relationship:
\begin{equation}
\label{2}t-t_0=T+\Delta_R(T)+\Delta_E(T)+\Delta_S(T)+\Delta_A(T)+%
\Delta_B(T).
\end{equation}
Here $t_0$ is the initial barycentric epoch of observation; $\Delta_R$, $%
\Delta_E$ and $\Delta_A$ correspond to the well-known Roemer, Einstein and
aberration time delays, whose exact mathematical expressions can be found in
the paper of Damour \& Taylor (1992); $\Delta_S$ is the Shapiro delay of
radio pulses in the gravitational field of companion of the pulsar (Epstein
1977; Damour \& Taylor 1992):
\begin{equation}
\label{3} \Delta_S=-2r\ln L,
\end{equation}
\begin{equation}
\label{4} L=1-e\cos u- s\{\sin\omega(\cos u-e)+(1-e^2)^{1/2}\cos\omega\sin u
\},
\end{equation}
which depends on the just introduced post-Keplerian parameters $r$ and $s$
as well as the classical orbital parameters: eccentric anomaly $u$,
eccentricity $e$, and longitude of the periastron $\omega $.

$\Delta _B$ in the equation (\ref{2}) is the additional bending delay caused
by the deflection of the pulsar beam in the gravitational field of its
companion. It has been calculated from the condition of emission of the
pulsar's pulse toward the observer (Kopeikin 1992). This condition is
expressed in the pulsar's proper reference frame by the equation:
\begin{equation}
\label{4a}(\vec N \cdot(\vec \Omega \wedge \vec b))=0,
\end{equation}
which defines a relation between the number of pulse, which has been emitted,
and the moment
of the emission reckoned in the pulsar proper time scale $T$. Here $\vec N$
denotes the spatial components of isotropic unit vector tangent to the
trajectory of the emitted pulse, and directed toward the terrestrial
observer; $\vec b$ is the unit vector in the fiduciary direction of
maximal pulsar radio emission (it is not necessarily the direction of
magnetic dipole); $\vec \Omega$ is the vector of the instantaneous angular
velocity of the pulsar; the signs ''$\cdot $'' and ''$\wedge $'' denote the
Euclidean scalar and cross products. We have assumed implicitly that the model
of the conical rotating beacon for the description of the pulsar's emission
(Lyne \& Manchester 1988) is true. However, we would like to point out that
calculations being described in this paragraph are still valid in
the case of a deviation of the polar beam geometry from the conical one as
it is, for instance, in the model of Narayan \& Vivekanand (1983).

Vector $\vec N$ does not keep a constant spatial orientation from the point of
view of an external observer at rest but has small (caused by both
classical and relativistic effects) spatial variations with respect to the
constant unit vector $\vec n$ referred to the rest frame of the binary
pulsar barycenter. To show this we  use the simplified version of the
relativistic relationship between the spatial components of the vectors $%
\vec N$ and $\vec n$ derived by Klioner \& Kopeikin (1992) in the case of
the relativistic N-body problem:
\begin{eqnarray}
\label{4b}\vec N&=&\vec n-\frac 1R(\vec n\wedge (\vec r_p\wedge \vec n))+
               \frac1c(\vec n\wedge (\vec v_p\wedge \vec n))
\nonumber \\
 & & - \frac{2Gm_c}{c^2r_R}\frac{(\vec n\wedge
         (\vec r_R\wedge \vec n))}{r_R+(\vec n\cdot \vec r_R)},
\end{eqnarray}
to replace in the equation (\ref{4a}) the time dependent components $\vec N$
by the ones of the vector $\vec n$. The complete relation between $\vec N$
and $\vec n$ (Klioner \& Kopeikin 1992) can be easily used for investigations
of the second order aberration effects and physical parameters of a
rotating black hole being considered as a possible companion of a pulsar
(Narayan {\it et al. }1991).

In formula ($\ref{4b})$, $c$ is the speed of light, $\vec r_p$ and $\vec
r_R$ are the radius-vectors pointing respectively from the barycenter of the
binary system and the center-of-mass of the pulsar's companion to the pulsar
(do not confuse these vectors with the notation $r$ for the
parameter characterizing amplitude of the Shapiro delay), $\vec v_p=d\vec
r_p/dT$ is the orbital velocity of the pulsar, and $R$ is the distance
between the binary pulsar and the solar system barycenters. Let us note that
it is possible to equate the vector $-\vec n$ to the unit vector $\vec K$
(that is $\vec n=-\vec K$) pointing from the Earth toward the pulsar system
as it is usually done, though such procedure requires an appropriate
justification from a theoretical point of view.

After substituting expression (6) into equation (5), expressing vectors $\vec
N$,
$\vec \Omega$, and $\vec n$ through the Euler variables, and solving the
obtained
trigonometric equation with respect to the fast Euler variable as described in
(Doroshenko {\it
et al. 1995} we get the counting formula (1) with additional phase corrections
due to
the classical and relativistic terms in the relationship (6). To be specific,
the second term in the right-hand side of this formula gives an additional but
negligibly small classical parallax delay in the rotational phase of
the pulsar (see, however, the paper of Kopeikin (1995) where the influence
of the orbital parallax effect on the propagation time of pulsar radio
signals is shown to be detectable); the third term in the right-hand side of
(6)
leads to the aberration delay $\Delta_A(T)$ depending on two directly
unobservable
parameters $A$ and $B$ introduced by Damour \& Deruelle (1986), and the very
last term defines after straightforward calculations the gravitational bending
delay
being under consideration:
\begin{equation}
\label{5}\Delta _B=L^{-1}\left( C\sin (\omega +A_e)+D\cos (\omega
+A_e)\right) .
\end{equation}
Here the true orbital anomaly $A_e$ is related to $u$ by the well-known
equation:
\begin{equation}
\label{8}A_e=2\arctan \left[ \left( \frac{1+e}{1-e}\right) ^{1/2}\tan \frac
u2\right] ,
\end{equation}
$a_R$ designates the semi-major axis of the relative orbit of the pulsar
with respect to its companion, and the constants $C$ and $D$ have been
introduced as the new post-Keplerian parameters:
\begin{equation}
\label{6}C=\frac{cr(1-s^2)^{1/2}}{\pi \nu _pa_R}\frac{\cos \eta }{\sin
\lambda },
\end{equation}
\begin{equation}
\label{7}D=-\frac{cr}{\pi \nu _pa_R}\frac{\sin \eta }{\sin \lambda }.
\end{equation}
where $a_R=a_p(m_p+m_c)/m_c$, and $a_p$ is the major semi-axis of the pulsar
orbit with respect to the barycenter of the binary system. It is interesting
to note that the final expression for the BD effect, formula (\ref{5}), does
not include any dependence on the spatial orientation of vector $\vec b.$
The same is true for the aberration delay $\Delta _A$.

For a better understanding of influence of the BD effect on the TOA of the
pulsar pulses let us compare the behavior of the functions $\Delta_S$ and $%
\Delta_B$ by inserting parameters of the system PSR B1855+09 into the
expressions (\ref{3}) and (\ref{5}). The explicit dependence on
time of the Shapiro (eq. (\ref{3})) and BD (eq. (\ref{5})) effects are shown
in Fig.1 , which has been drawn using observational parameters of the PSR
B1855+09 reported by Kaspi {\em et al. }(1994).

\begin{figure}
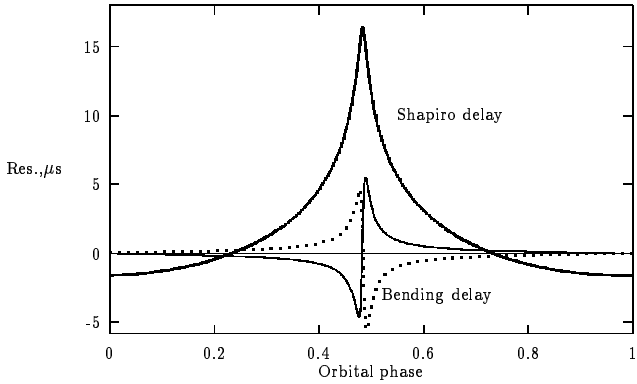

\vspace{5.cm}
\caption{
    The illustrative example of the TOA residuals for PSR
    B1855+09 caused by the Shapiro delay and deflection of
    the pulsar beam in the gravitational field of its
    companion - smooth or dotted curves. For more
    explanations see the text.
          }
\end{figure}

For illustrative purposes we have chosen the artificially small numerical value
for $\lambda =0^{\circ }.005,$ and two different values for the angle: $%
\eta=-85^{\circ}$ (solid curve, $D>0,C>0$), and $\eta=85^{\circ }$ (dotted
curve, $D<0,C>0$) to depict the BD effect in the most clear
form. One can see from Fig.1 that the modulation of TOA due to the BD effect is
completely different and quite sharper than that caused by the Shapiro
delay. The total observational timing effect will be the sum of the Shapiro
and BD delay curves. It is important to note that the asymmetry in the curve
describing the BD effect does not depend on the value of the angle $\lambda $
(since $0\leq $$\sin \lambda \leq 1,$ when $0^{\circ }\leq \lambda \leq
180^{\circ }$) and is sensitive to the numerical value of the angle $\eta $
only. Therefore, we can say nothing from the investigation of the BD effect
whether the pulsar spin axis $\vec S$ is along the direction to observer or
in the opposite one.

The considered example shows that there is in principal a possibility to detect
the BD effect in the binary pulsars with nearly edgewise orbits. However,
this effect was never included in existing pulsar timing algorithms.
Therefore, it is important to check how strongly the BD effect perturbs
measured parameters of such binary systems.

\section{ Measurability of masses and orbital inclination in the edge-on
binary systems}

To investigate the measurability of the BD effect we shall
follow the linearization procedure for evaluation of the pulsar parameters,
developed by Blandford \& Teukolsky (1976), and improved by Damour \&
Deruelle (1986). The procedure is based upon the, so called, ''periodic
approximation'' of differential timing formula (see below). The idea is to
introduce instead of the total set of orbital and spin parameters of the
pulsar in the timing algorithm a convenient set of independent parameters,
which absorb all the secular time dependencies in the form of polynomials of
time $t$. Coefficients of the differential timing formula are in such case
only periodic functions of time.

Let the observed rotational phase of the pulsar be $N_{obs}$. Then we can
improve the initial estimates of any pulsar's parameter $e_i^{(0)}$ using
the residuals of the pulsar's rotational phase $%
R(t_k)=N_{obs}(t_k)-N_{cal}(e_i^{(0)},t_k)$, where the numerical values of $%
N_{cal}$ are computed at the moments of observed arrival times $t_k$ with
the help of the equations (\ref{1}) - (\ref{4}) and (\ref{5}) - (\ref
{7}). Assuming that the differences $\delta e_i=e_i-e_i^{(0)}$ between the
(yet unknown) true and initial numerical values of the parameters are small,
one can represent an expression for the total derivative of the phase
residuals (the differential timing formula) in the form:
\begin{equation}
\label{9}R(t)=\sum_{i=1}^7B_i\cdot \delta e_i\quad ,
\end{equation}
where the variations of the adjusted parameters in formula (\ref{9}) are:
\begin{eqnarray}
\delta e_1&=&-\frac{1}{\nu_p}(\delta N+t\cdot\delta\nu_p+
   \frac{t^2}{2}\cdot\delta\dot\nu_p+\cdots)
\nonumber \\
\delta e_2&=&\delta \alpha
\nonumber \\
\delta e_3&=&\delta(\beta+\gamma)
\nonumber \\
\label{11}\delta e_4&=&-\delta\sigma-t\cdot\delta n\\
\delta e_5&=&\delta e
\nonumber \\
\delta e_{\mu}&=&\frac{\delta m_{\rm c}}{m_{\rm c}}
\nonumber \\
\delta e_{\xi}&=&\frac{\delta\sin i}{\sin i} ,
\nonumber \\
\nonumber
\end{eqnarray}
with $\alpha = x\sin \omega $ and $\beta = \sqrt{1-e^2}\ x\cos \omega $; $%
\sigma $ is the orbital reference phase; $n\equiv 2\pi /P_b$, $P_b$ is the
orbital period; $\gamma $ is the gravitational red shift plus transverse
Doppler effect. Let us point out that to get relative accuracies in the
determination of the mass $m_c$ and sin$i$ we have used the substitutions for
the parameters of the type $m_c=m_c^{(0)}\cdot (1+\mu )$ and $\sin i=\sin
i^{(0)}\cdot (1+\xi )$, where $m_c^{(0)}$ and $\sin i^{(0)}$ are supposed to
be fixed and equal to the current estimated values for these parameters.

The functions $B_i=\partial N(t)/{\partial e_i}$ are the partial derivatives
of the pulsar rotational phase $N(t)$ with respect to the $e_i$-th adjusted
parameter of the pulsar, and we neglect the relativistic post-Keplerian
parameters $\delta _r$, $\delta _\theta $, $k$ and all hidden
parameters (see for
explanation of the physical meaning of the omited parameters the papers of
Damour \& Deruelle (1986) and Kopeikin (1994)):
\begin{equation}
\label{12}B_1=1,
\end{equation}
\begin{equation}
\label{13}B_2=\cos u-e,
\end{equation}
\begin{equation}
\label{14}B_3=\sin u,
\end{equation}
\begin{equation}
\label{15}B_4=\frac{\alpha \sin u-\beta \cos u}{1-e\cos u},
\end{equation}
\begin{equation}
\label{16}B_5=-\alpha -B_4\sin u,
\end{equation}
\begin{equation}
\label{17}B_6\equiv B_\mu =\Delta_S+\Delta_B,
\end{equation}
\begin{eqnarray}
\label{18} B_7\equiv B_\xi=&L^{-1}s&
            \left[(2r+\Delta_B)(1-e\cos u)-\frac{sC}{1-s^2}\right]
\nonumber \\
  & &  \times \sin(\omega+A_e).
\end{eqnarray}
After the first step of the least-squares fit of the residuals $R(t)$, the
obtained corrections for the pulsar parameters $\delta e_i^{(0)}$ are added
to their initial values $e_i^{(1)}=e_i^{(0)}+\delta e_i^{(0)}$ and are used
again at the second step of the iteration procedure. The process is repeated
untill the limited values of the estimated parameters are obtained.

A similar model has been used by Ryba \& Taylor (1991) and Kaspi {\em et al.}
(1994) in the measurement of the parameters of the PSR B1855+09. However,
these authors have relied upon the algorithm developed by Damour \& Deruelle
(1986). The main difference between our
differential timing formula and the one
of Damour \& Deruelle (1986) is in the expressions for the partial
derivatives $B_6$ and $B_7$ taking into account the BD effect in the
explicit form. It is important to note that omitting the BD effect
from the timing algorithm will result, in principle, in incorrect estimates
for the relativistic post-Keplerian parameters $r$ and $s.$ This fact
must be take into account for unambigious testing of the alternative gravity
theories in the strong field regime.

Using our improved expressions (\ref{17}) and (\ref{18}) for these functions
one can examine how the BD effect will manifest itself in the fitted values
of the pulsar parameters $\mu $ and $\xi .$ To investigate this question we
take numerical simulations
using as an example imaginary binary systems having the same parameters
as the pulsars B1855+09 and\ B1913+16 exept for the inclination angle $i$ of
the orbit. The inclination is varied in the range from 86.5$^{\circ }$ to 90$%
^{\circ }$. Let us suppose that the standard deviation of the mock arrival
times is $\epsilon $ (a constant measured in $\mu $sec), and covariance
matrix $C_{ij}$ for the fitting method is $\epsilon ^2$
times the inverse of the 7$\times $7 matrix:
\begin{equation}
\label{19}B_{ij}=\sum_{k=1}^nB_i(t_k)\cdot B_j(t_k).
\end{equation}
Here $n$ is the total number of the TOA measurements and the functions
$B_i$ $(i = 1,2,...7)$ are given by equations (\ref{12}) - (\ref{18}).
Then our variance estimates for the evaluated parameters $e_i$ are:

\begin{equation}
\label{21}\delta e_i=\sigma _{e_i}\cdot \frac{\epsilon (\mu s)}{\sqrt{n}}%
\quad \quad ,
\end{equation}
where the errors $\sigma _{e_i}=\sqrt{b_{ij}}$ of the parameters are
computed by using the diagonal elements of the inverse matrix $%
b_{ij}=B_{ij}^{-1}.$ We also suppose in our numerical simulations that the
observational points are equally distributed over the pulsar's orbital phase
and their total number is equal to 1000. Two different timing models have
been compared, namely, one of Damour \& Deruelle (1986) without the BD
effect, and that outlined above, which includes the BD effect.

First of all, we took calculations for binary systems having orbital parameters
like the PSR B1855+09. Figs 2 and 3
illustrate the behavior of the fractional uncertainties $\sigma_\mu $ = $\sigma
_{m_c}/m_c$ and $\sigma_\xi $ = $\sigma_{\sin i}/\sin i$ plotted as
functions of $\cos i$ for these two algorithms respectively,

\begin{figure}
\vspace{5.5cm}
 \caption{
    The values of log$_{10}(\sigma_\mu )= $log$_{10}$($\sigma_{
    {\rm m_c}}/{\rm m_c}$) vs. ($\cos i$) for measurement of the
    pulsar's companion mass in the two timing models for binary
    pulsars like PSR\ B1855+09. The dotted curve (S) represents
    the values $\sigma _\mu $ being expected from the data fit
    according to the Damour \& Deruelle (1986) timing algorithm,
    and the two solid curves represent the values $\sigma _\mu $
    being expected from the data fit with the gravitational
    bending delay taken into account. The assumed values for the
    angles $\eta =-85^{\circ };$ $\lambda =0.1^{\circ },$ and
    $1^{\circ }$ correspond to the curves labeled as $(0.1),$ and
    $(1)$ respectively.  The long tic on the bottom horizontal
    axis denotes the value of $\sin i=0.9992$ presented in (Kaspi
    {\em et al. }1994).
          }
 \end{figure}

\begin{figure}
\vspace{5.5cm}
 \caption{
    The same as Fig.2, except for only that the plotted
    values are log$_{10}(\sigma _\xi )=$log$_{10}$($\sigma _{\sin
    i}/\sin i$) vs. ($\cos i$ ).
          }
 \end{figure}

\noindent
where we fixed the value $\eta =-85^{\circ },$ and supposed the inclination
angle $i$ ranges from 86.5$^{\circ }$ to 90$^{\circ }$. Our assumed values
for $\lambda =0^{\circ }.1,$ and $1^{\circ }$ were chosen for illustrative
purposes. They correspond for the case of the PSR B1855+09 system to the values
of the parameter $D=-18ns,$ and $-1.8ns$, respectively. Note that the numerical
values for the parameter $C$ depend on the inclination angle, and have been
computed from the relation $C=-D\cot \eta \cos i$ following from equations (
\ref{6}) and (\ref{7}). This relation between $C$ and $D$ has been used in
equations (\ref{17}), (\ref{18}) for drawing the plots. In fact, the values
of $C$ for the chosen set of $D$ and $\sin i$ lie in the range $0$ $%
\div 0.1ns$. Replacing the positive sign of the parameter $D$ with the
negative one with the same numerical values (that corresponds to the
positive value of angle $\eta =85^{\circ }$) does not change the Figures 2
and 3 perceptibly.

The same procedure has been applied to estimate the measurability of the
''range'' and ''shape'' parameters in the case of binary system PSR\
B1913+16. In such systems, with more massive pulsar companions, one
can expect to measure the BD effect more easily. We took four
numerical values for $\lambda =0^{\circ
}.1, 1^{\circ },$ $10^{\circ },$ and $80^{\circ },$ that corresponds to $%
D=-15721ns,-1572ns,-158ns,$ and $-28ns.$ The parameter $C$ lies in the
range $0\div 0.2ns. $ Results of the calculations are shown in the
figures 4 and 5.

\begin{figure}
\vspace{5.5cm}
 \caption{
    The values of log$_{10}(\sigma_\mu )=$log$_{10}$($\sigma_{
    {\rm m_c}}/{\rm m_c}$) vs. ($\cos i$) for measurement of the
    pulsar's companion mass in the two timing models for binary
    pulsars like PSR\ B1913+16. The dotted curve (S) represents
    the values $\sigma _\mu $ being expected from the data fit
    according to the Damour \& Deruelle (1986) timing algorithm,
    and the four solid curves represent the values $\sigma _\mu $
    being expected from the data fit with the gravitational
    bending delay taken into account. The assumed values for the
    angles $\eta =-85^{\circ };$ $ \lambda =0^{\circ }.1,1^{\circ
    },10^{\circ },$ and $80^{\circ }$ correspond to the curves
    labeled as $(0.1),\quad (1),$ $(10),$and $(80)$ respectively.
    Longitude of the periastron at the initial epoch $\omega
    _0=180^{\circ }.$
          }
 \end{figure}

\begin{figure}
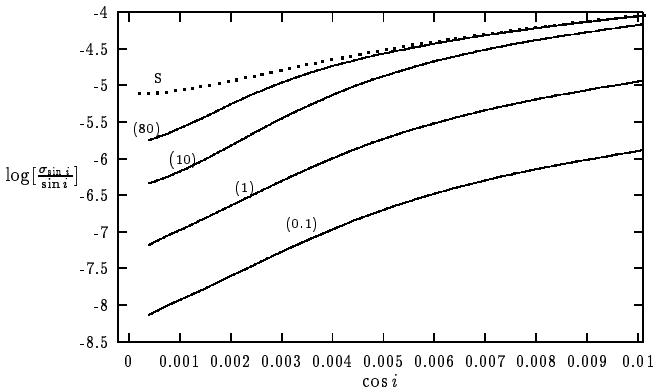

\vspace{5.5cm}
 \caption{
   The same as Fig. 4, except for only that the plotted values
   are log$_{10}(\sigma_\xi )=log_{10}(\sigma_{\sin i}/\sin i)$
   vs. ($\cos i$ ).
          }
 \end{figure}

As one can see from the plots (2) and (3), the inclusion of the effect of
the gravitational bending delay is of no practical consequence for
reducing the relative errors in the mass and $\sin i$ determinations in such
systems as the binary pulsars like PSR B1855+09 when $\cos i>0.01$ and $%
\lambda >1^{\circ }$. As for the binary pulsars like PSR\ B1913+16,
possessing more massive pulsar's companion, the BD effect has no influence
when $\cos i>0.01$ and $\lambda >80^{\circ }$ . From an evolutionary point of
view it is expected that for the pulsars under consideration $\lambda \simeq
i\simeq 90^{\circ }$ and the only real hope to measure the influence of
the BD effect
is in this case when $\cos i<0.003.$ This value is close enough to
that anticipated by Epstein (1977). It should be noted, however, that these
estimates are true only for the total number of ''observational points''
being equal to 1000. Taking into account many more points or concentrating
observations near the moment of superior conjunction makes the obtained
estimates less conservative.

\section{ Measurability of the gravitational deflection of light parameters
in the PSR B1855+09}

The numerical simulations made in the previous paragraph
show that the amplitude of
the BD effect is, probably, too small to be detected with confidence in the
PSR\ B1855+09 since the real value of $\cos i = 0.04$. Nevertheless, in the
same spirit of ''looking for zero to be sure it is there'', we have tried to
find it using the 7-yrs. observational data set for PSR B1855+09
obtained by J. Taylor and collaborators at the Arecibo observatory (Kaspi
{\em et al.,} 1994). To process the observational data we apply the timing
program TIMAPR described in (Doroshenko \& Kopeikin 1990). It has been
generalized for timing of binary pulsars in accordance with the timing
algorithm described in Sec. 3 with a secular drift of all relevant parameters.
The celestial reference frame was based upon the DE200/LE200 ephemerides of the
Jet Propulsion Laboratory (Standish 1982) and
the relativistic time scales transformations in the Solar system developed
by Fairhead \& Bretagnon (1989) are used
in our fitting procedure (see the paper of Brumberg \& Kopeikin (1990) for more
theoretical details concerning the origin and structure
of these transformation).
The polar motion corrections were introduced using
the ''raw values of X,Y, universal time for every 5 days'' refered
to the 1979 BIH system
(BIH Annual Report) and the ''normal values of the Earth orientation
parameters at 5-day intervals'' refered to the EOP (IERS)
C02 system (IERS Annual Report).

The general concept of measurability of pulsar parameters is described in the
papers of Taylor \& Weisberg (1989) and Damour \& Taylor (1992), and we
follow the outlined procedure considering now $C$ and $D$ in the formula (%
\ref{5}) as the new independent fitted parameters$.$ It changes the formula (%
\ref{9}) to:

\begin{equation}
\label{25}R(t)=\sum_{i=1}^9\hat B_i\cdot \delta e_i\quad ,
\end{equation}
where parameters $e_1,...,e_5$ are the same as in the equations (14), $%
\{e_i;6\leq i\leq 9\}=\{r,s,C,D\}.$ Functions $\hat B_i=B_i$ for $%
i=1,2,3,4,5 $ are taken from the equations (\ref{12})-(\ref{16}) and:
\begin{equation}
\label{26}\hat B_6\equiv \hat B_r=-2\ln L,
\end{equation}
\begin{equation}
\label{27}\hat B_7\equiv \hat B_s=L^{-1}(2r+\Delta _B)(1-e\cos u)\sin
(\omega +A_e),
\end{equation}
\begin{equation}
\label{28}\hat B_8\equiv \hat B_C=L^{-1}\sin (\omega +A_e),
\end{equation}
\begin{equation}
\label{29}\hat B_9\equiv \hat B_D=L^{-1}\cos (\omega +A_e).
\end{equation}

To determine parameters $r,s,C$ and $D$ we used the least-squares procedure
of minimizing the statistic:
\begin{equation}
\chi ^2=\sum_{i=1}^n\Bigl[\frac{N(t_i)-N_i}{\sigma _i\nu _p}\Bigr]^2,
\end{equation}
where $\sigma_i$ is the estimated uncertainty of the TOA $t_i.$ The
statistic $\chi^2$ measures the goodness of fit of the timing model. However,
direct application of the least-squares method with all parameters
considered to be varied freely fails to determine the parameters $C$ and $D$
because of their smallness. Moreover, the parameter $C$ includes the
additional, small factor $\cos i=0.04$, and, for this reason, has been
assumed to be equal to zero identically. Then, to extract
some valuable information,
we investigate the structure
of the function $\Delta \chi ^2(e_i)=\chi ^2(e_i)-\chi _0^2(e_i^{est})$ in the
neighborhood of the global minimum $\chi _0^2$ in the multi-dimensional
space of the fitted parameters (Taylor \& Weisberg ,1989; Ryba and Taylor
,1991; Damour \& Taylor ,1992).

We have calculated the functional dependence of $\chi ^2$ statistic near the
region of its global minimum with the minimization of the $\chi ^2$ by the
fit of the astrometric $\{N_0,\alpha ,\delta ,\mu _\alpha ,\mu _\delta \}$,
spin $\{\nu _p,\dot \nu _p,\ddot \nu _p\}$, orbital Keplerian $\{x=a\sin
i,e,T_0,\omega ,P_b\},$ and post-Keplerian $\{\dot P_b,\dot e,\dot \omega
,\} $ parameters of the pulsar with the parallax $\pi $, dispersion measure $%
DM$ , and post-Keplerian parameters $r$ and $s$ held fixed at the unique
values reported by Kaspi {\em et al.} (1994). In this procedure the
parameter $D$ was held fixed at a number of different points from -40 ns to 40
ns. The results of the fitting procedure are shown in Fig. 6.

\begin{figure}
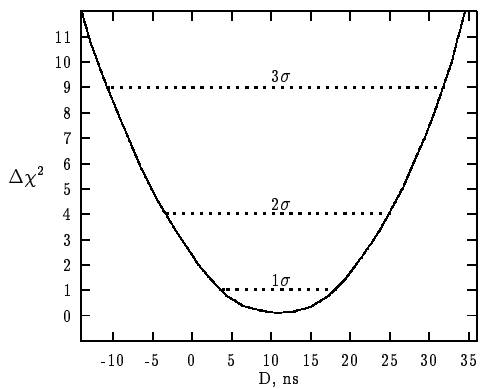

\vspace{5.5cm}
 \caption{
    The smooth curve represents values of $\Delta\chi^2$
    obtained in data fit with the fixed values of the $r$ and $s$
    relativistic post-Keplerian parameters obtained by Kaspi {\em
    et al.} (1994). The levels of confidence, corresponding to
    1$\sigma$, 2$\sigma$ and 3$\sigma $ determination of
    parameter $D$ are plotted with dotted lines.
         }
  \end{figure}

As one can see from this plot the corresponding curve has a parabolic shape,
and the numerical value of the parameter $D=12_{-9}^{+10}$ nanosecond is not
zero with the 68 \% confidence ( 1$\sigma$ ) limits. The values of other
astrometric, spin, Keplerian, and post-Keplerian parameters for the PSR
B1855+09 are in a perfect ( 1$\sigma$ ) agreement with the values reported
by Kaspi {\em et al.} (1994). Although we can not say anything with
the adopted
( 3$\sigma$ ) confidence about the real numerical value of the BD effect, the
parabolic structure of the $\chi^2$ near region of its global minimum (with
respect to the parameter $D$) indicates that the effect in question can be,
perhaps, measured in the future.

\section{Conclusion}

We have derived the improved timing formula (\ref{2}) for binary pulsars
taking into account the relativistic effect of light deflection in the
gravitational field of the pulsar's companion. It has been shown the effect
is perceptible only for the pulsars with nearly edge-on orbits. The measurement
of the effect in question permits, in principle, the determination of the
position angle $\eta $ of the angular velocity vector of the pulsar although
it fails to determine direction of the rotation. The influence
of the gravitational
deflection of light on the measurability of the magnitude $r$ and shape $s$
parameters of the Shapiro time delay as well as the BD parameters $C$ and $D$
has been examined. Processing of the observational data for the binary
pulsar B1855+09 was rendered in an effort to find some
indication on the presence
of effect of the gravitational deflection of light in this system.

Taking into account the effect of the gravitational deflection of light in the
edge-on binary pulsars allows us to make more precise experimental tests of
alternative theories of gravity in the strong gravitational field regime
when the cosine of the orbital inclination to the plane of the sky is less
than 0.003. This rather conservative estimate can be relaxed if
the rotational axis of the pulsar is close enough to the line of sight and/or
the total number of equally spaced observations exceeds 1000.

After this paper was submitted for publication, we were
 informed about a paper by
Goicoechea {\it et al.} where the BD effect is discussed for the case of
binary pulsars and X-ray binaries on circular orbits.

\subsection*{Acknowledgments}

We are very grateful to Kaspi V.M., Taylor J.H. and Ryba M.F. for their kind
permission to use their observational data on the PSR\ B1855+09 binary
system before publication. We would also like to acknowledge K.Yokoyama (NAO
of Japan, Mizusawa) who supplied us the polar motion corrections in machine
readable form. We are thankful Yu.P.Ilyasov, K.Yokoyama, T.Fukushima, and
H.Kinoshita for participation in discussions. E.L.Presman made valuable
remarks concerning statistical significance of the obtained results and we
highly appreciate him. An expression of gratitude is due to Dr M.Holman for
polishing the grammar of the manuscript and anonimous referee for helpful
comments and suggestions.

S.M.Kopeikin is deeply indebted to the staff of the National Astronomical
Observatory of Japan (Mitaka) for the support of this work. O.Doroshenko
has been partially supported by the ISF grant M99000.

\label{lastpage}

\end{document}